\documentclass[11pt,a4paper,reqno, aps, superscriptaddress, 
amssymb,amsmath,amsfonts]{amsart}

\usepackage{amsopn}
\usepackage{amsmath}
\usepackage{amssymb}
\usepackage{amsthm}
\usepackage{hyperref}
\usepackage{graphicx}
\usepackage{color}
\usepackage{listings}
\usepackage{mathtools}
\usepackage{subcaption}
\usepackage{enumerate}
\usepackage{cite}
\usepackage{todonotes}
\usepackage[foot]{amsaddr}
\usepackage{amsfonts}

\usepackage{multirow}
\usepackage{adjustbox}
\usepackage{geometry}
\usepackage{tikz}
\usetikzlibrary{angles,quotes}
\usetikzlibrary{decorations.pathreplacing}
\DeclareMathOperator{\tr}{tr}

\newcommand{\R}{\ensuremath{\mathbb{R}}}

\newcommand{\C}{\ensuremath{\mathbb{C}}}

\newcommand{\ket}[1]{\ensuremath{|#1\rangle}}
\newcommand{\bra}[1]{\ensuremath{\langle#1|}}
\newcommand{\ketbra}[2]{\ensuremath{\ket{#1} \! \bra{#2}}}
\newcommand{\proj}[1]{\ensuremath{\ketbra{#1}{#1}}}

\newcommand{\Id}{{\rm 1\hspace{-0.9mm}l}}

\newcommand{\QQ}{\mathcal{Q}}

\newcommand{\EE}{\mathcal{E}}
\newcommand{\XX}{\mathcal{X}}

\newcommand{\YY}{\mathcal{Y}}

\newcommand{\OO}{\mathcal{O}}

\newcommand{\CC}{\mathcal{C}}

\renewcommand{\SS}{\mathcal{S}}

\newcommand{\A}{\mathcal{A}}
\newcommand{\B}{\mathcal{B}}

\newcommand{\inner}[2]{\ensuremath{\langle{#1} \,,{#2}\rangle}}
\newcommand{\ketV}[1]{\ensuremath{|#1\rangle\!\rangle}}
\newcommand{\braV}[1]{\ensuremath{\langle\!\langle#1|}}
\newcommand{\ketbraV}[2]{\ensuremath{\ketV{#1}\braV{#2}}}
\newcommand{\projV}[1]{\ensuremath{\ketbraV{#1}{#1}}}

\newtheorem*{rem*}{Remark}

\usepackage{caption}
\def\>{\rangle}
\def\<{\langle}
\title{Shared entanglement 
	for three-party causal order guessing game}

\author{Ryszard Kukulski$^{1}$, Paulina Lewandowska*$^{2}$, \and Karol 
Życzkowski$^{1,3}$}
\address{$^1$ Faculty of Physics, Astronomy and Applied Computer Science, Jagiellonian University, ul. Łojasiewicza 11, 30-348 Kraków,
	Poland}
\address{$^2$ IT4Innovations, VSB~-~Technical University of Ostrava, 17.~listopadu 2172/15, 708 33 Ostrava, Czech Republic}
\address{$^3$ Center for Theoretical Physics, Polish Academy of Sciences, Al. Lotników 32/46, 02-668 Warszawa, Poland}

\begin{document}
\maketitle

\begin{abstract}
In a variant of communication tasks, players cooperate in choosing their local strategies to compute a given task later, working separately. Utilizing quantum bits for communication and sharing entanglement between 
parties is a recognized method to enhance performance in these situations. In 
this work, we introduce the game for which three parties, Alice, Bob and 
Charlie, would like to discover the hidden order in which they make the moves. 
We 
show the advantage of quantum strategies that use shared entanglement and 
local operations over classical setups for discriminating operations' 
composition order. The role of quantum resources improving the 
probability of successful discrimination is also investigated. 
\end{abstract}

\section{Introduction}
The theory of resources \cite{coecke2016mathematical}, in various domains such as economics, information theory, and physics, revolve around 
allocating, utilizing, and managing assets or elements that are valuable or essential for achieving specific goals or tasks. In information theory 
\cite{watrous2018theory, nielsen2010quantum}, we investigate how these 
resources can effectively transmit, process, and store information. It 
explores concepts such as data compression, error correction, and channel 
capacity, aiming to optimize limited resources to achieve desired 
communication tasks.

In a communication task, there are usually multiple parties, often referred to as players or agents, each having some input data relevant to the problem. 
Players cooperate in choosing their local strategies to compute a given goal 
later, working separately. The task is to compute a desired function of the 
combined inputs while minimizing the amount of information 
exchanged between the parties. Communication efficiency is typically measured by the number of bits or messages exchanged between parties. This 
paper studies an explicit communication task for which Alice, Bob and Charlie 
meet to discuss a strategy for which, after separation, they would like to 
discover the hidden order in which their moves are made.

  It is well known that quantum resources offer a significant advantage over 
  classical ones 
  \cite{chitambar2019quantum,guerin2016exponential,brassard2003quantum, 
  brukner2004bell, schmid2020type}. Utilizing quantum bits for communication 
  and sharing entanglement \cite{hsieh2010entanglement} between parties is 
  one of the recognized methods to enhance the likelihood of success in 
  communication scenarios. Such procedures are called shared entanglement and 
  local 
  operations (LOSE) strategies \cite{bennett1996concentrating}. 
  
 In the communication scenario considered in this paper, Alice, Bob and 
 Charlie are guessing the hidden order in which they are making moves. 
 Their priority is to achieve the highest possible probability of a correct 
 guess. We analyze the value of this probability depending on the type of 
 strategies all parties may utilize. The following strategies are considered: 
 classical memoryless strategy, quantum memoryless strategy, classical 
 strategy based on local operations and shared randomness, classical 
 non-signaling strategy and quantum strategy based on local operations and 
 shared entanglement. For memoryless cases, we prove that the optimal 
 probability of discrimination order equals $1/3$. For extended 
 versions of classical strategy, we prove that either local operations and 
 shared randomness or non-signaling strategies provides the probability 
 $5/6$. Finally, we show the advantage of using quantum LOSE strategy, 
 for which we can achieve perfect discrimination. 
 
This paper is organized as follows.   Section \ref{sec:scenario} presents the concept of the communication scenario and trivial examples. In Section \ref{sec:preliminary} we introduce 
necessary mathematical
framework. Section \ref{sec:main} is dedicated to presenting the main results 
of this work. First, in Subsection \ref{classical-basic}, we consider 
classical memoryless strategy. Next, in Subsection \ref{losr} and Subsection 
\ref{nonsign}, we present extended approaches to the classical scenario based 
on local operations and shared randomness and non-signalling operations. 
Subsection \ref{quantum-basic}, whereas shows the quantum version of the 
memoryless scenario. Finally, in Subsection \ref{lose}, we use quantum 
entanglement to achieve perfect discrimination of order guessing. 
Concluding remarks  are presented 
in the final Section \ref{sec:conclusion}.

\section{Causal order guessing game}\label{sec:scenario}
{\bf \textit{Scenario:}}
Let us imagine that we have a game of guessing order between three parties 
involved: Alice, Bob, and Charlie. They are given the following instructions. 
All of them will be randomly ordered in a line, but the choice of the order 
will not be known to them. The first person in the line will obtain a bit 
(qubit), which is a given initial state of the system. Then, he can send an 
arbitrary bit (qubit) of information to the second person in the line, the 
second one to the third one and the third person's bit (qubit) is the output 
of the system. By investigating the output of this simple communication 
procedure, Alice, Bob and Charlie should be able to guess, which order they 
were put in. They are allowed to cooperate at the beginning of the game to 
choose their strategy and prepare the initial state (bit or qubit) of the 
system. They are allowed also to use personal auxiliary systems to write down 
some important information, for example, the value of the system they received.
Moreover, after the communication stage, they may consult the results, 
discuss the states they sent or obtained, or inspect the state of the system. 
The joint decision will indicate one of six possible orders they could be. 
They win the guessing game if they correctly guess the order. The scope of our 
interest is to find a strategy that maximizes the probability of correct 
guessing.\\

{\bf \textit{Two-party case:}} Let us start simply with only Alice and Bob 
playing this game. In that case, they can always successfully guess the order 
by utilizing the following strategy. They prepare the initial classical state 
as 
$\rho = 0$. Alice communication strategy is to always return the state $1$, that is $a(0) = 1$ and $a(1) = 1$. Bob's strategy is to negate the input he 
received, that is, $b(0) = 1$ and $b(1) = 0$. If Alice is first, we have $ba(0) = b(1) = 0$. Otherwise, if Bob is first, we get $ab(0) = a(1) = 1$. 
After the communication round, Alice and Bob can simply read the system's state to correctly determine the order.\\

{\bf \textit{Three-party case with a trit of information:}}  As the final 
warm-up, let us analyze the situation when Alice, Bob and Charlie play the 
game, but they may communicate by sharing trits (qutrits). To win this game, 
each of them can use the same strategy: remember the input state and return 
the value of the input state plus one in the modular arithmetic. The system is 
put into the state $\rho = 0$. The first person registers $0$ and returns $1$. 
The second person registers $1$ and returns $2$, while the last person 
registers $2$ and returns $0$. Each of the persons know precisely their 
position, and hence, they may determine the order ideally each time. 

{\bf \textit{Three-party case with a bit of information:}} The main result 
of this work consists of the analysis of the smallest, 
non-trivial case of the introduced causal order guessing game: three-party 
game, but with one unit (bit or qubit) of information. We present complete 
analysis of this case in Section~\ref{sec:main}.

\section{Notation and mathematical preliminaries}\label{sec:preliminary}

\subsection{Classical}\label{sec:preliminary-c}

Let us introduce the following notation for classical case. We consider real 
Euclidean spaces $\XX = \R^n$ and probability vectors $\rho$ defined on them. 
We will use the notation $\rho \in \Omega(\XX)$ to define a probability vector 
on $\XX$. Whenever $\rho$ will represent a measure concentrated on $i$-th 
coordinate, we will simply write $\rho = i$. For example, if $\rho = 
\begin{bmatrix}1\\0\end{bmatrix} \in \Omega(\R^2)$, we write $\rho = 0$ and 
for $\rho = \begin{bmatrix}0\\1\end{bmatrix}$ we use shortcut $\rho = 1$. By  
$\mathrm{L}(\XX, \YY)$, we denote 
the collection of all stochastic matrices of the form $M: \XX \rightarrow 
\YY$. Additionally, we will use shortcut $\mathrm{L}(\XX) = \mathrm{L}(\XX, 
\XX)$. A measurement of the classical state $\rho$ is given by a collection 
$\{\Omega_i\}_i \subset \XX$, such that the components of the vector $\sum_i 
\Omega_i$ are all equal $1$. Probability of registering $i$-th measurement 
effect is equal to inner product $\inner{\Omega_i}{\rho}$.

In this work, each of the parties, Alice, Bob and Charlie, will apply their 
local operation in the process of order guessing. We denote by $\SS = \R^2$ 
the shared bit space that parties use for communication. For Alice, we will 
denote her local stochastic operations by letters $a$. Bob and Charlie will be 
represented by $b$ and $c$, respectively. The unknown order of their 
operations will be indicated by permutations of symmetric group 
$S_3$~\cite{cameron1999permutation}. There are $3!$ permutations of that type, 
and we will use the notation of $\pi \in S_3$ in the article.
For a given permutation $\pi$ and operations $a,b,c$ by $\pi(a, b, c)$, we 
denote a composition of $a,b,c$ according to the order defined by $\pi$. 
The main aim of this work is to find an optimal strategy which maximizes 
the probability of a correct discrimination of order composition of these $3$ 
mappings. We assume that each order is equally likely
to occur, with the probability $1/3!$. The choice of the input 
state $\rho$ and the final measurement 
$\{Q_\pi\}_\pi$ is arbitrary. Mathematically, the objective function
can be written as  
\begin{equation}\label{eq-prob-def-c}
	p_c = \max_{\rho, a, b, c, \{\QQ_\pi\}_\pi} 
	\,\, \frac{1}{6} \sum_{\pi \in S_3} \inner{Q_\pi}{\pi(a,b,c)(\rho) }.
\end{equation}

\subsection{Quantum}\label{sec:preliminary-q}
In the quantum case, we will use the following notation. Complex Euclidean 
spaces we denote by $\XX = \C^n$. For a matrix $M: \XX \to \YY$ we use the 
notation $\ketV{M} \in \YY \otimes \XX$ to define the vectorization of the 
form $\ketV{M} = (M \otimes \Id)\sum_{i} \ket{i,i}$. By  $\mathrm{Pos}(\XX)$, 
we denote the set 
of  positive semidefinite operators acting on the vectors from $\XX$, whereas 
the set of quantum states, that is, positive semidefinite operators $\rho$ 
with unit trace, $\tr(\rho) = 1$, will be denoted by $\Omega(\XX)$. By  
$\mathrm{L}(\XX, \YY)$, we denote 
the collection of all quantum channels that acts on the operators defined on 
$\XX$, which return as outputs operators acting on $\YY$. The shortcut 
$\mathrm{L}(\XX)$ is reserved for quantum channels of the form 
$\mathrm{L}(\XX, \XX)$. A general quantum measurement that
is a positive operator valued measurement (POVM) is a 
collection of positive semidefinite
operators $\{ Q_i \}_i \subset \mathrm{Pos}(\XX)$ called effects, which sum up 
to identity,  $\sum_i Q_i = \Id$. According to the Born rule, the 
probability of obtaining $i$-th effect for a given quantum state $\rho$ 
equals $\tr(Q_i \rho)$.

Similarly as in the classical case, Alice, Bob and Charlie, will apply their 
local operation in the process of order guessing. Here, the shared qubit 
system will be $\SS = \C^2$ and the quantum operations will be denoted by $A, 
B, C$. For a given permutation $\pi \in S_3$ and operations $A, B, C$ by 
$\pi(A, B, C)$, we 
denote a composition of $A, B, C$ according to the order defined by $\pi$. 
In the quantum case, the maximal probability of correct discrimination of 
composition order is given by
\begin{equation}\label{eq-prob-def}
	p_q = \max_{\rho, A, B, C, \{\QQ_\pi\}_\pi} 
	\,\, \frac{1}{6} \sum_{\pi \in S_3} \tr \left( Q_\pi \pi(A, B, C)(\rho) 
	\right).
\end{equation}

\section{Main results}\label{sec:main}
We start our analysis with the classical case.

\subsection{\bf Classical memoryless strategy:}\label{classical-basic}
\begin{figure}[htp!]
	\includegraphics[scale=1.5, angle=0]{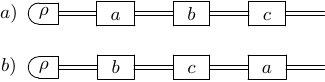}
	\caption{A graphical representation of possible orders of  mappings $a, b, 
	c$ acting on the initial 
	state $\rho$. Double lines represent a classical information. Time flows 
	from left to right. The  order of operations  a) $a \to b \to c$; b) $b 
	\to  c \to 
	a$.}
	\label{fig:cb}
\end{figure}

In this scenario Alice, Bob and Charlie are using classical, mono-partite 
stochastic maps $a, b, c \in \mathrm{L}(\SS)$, respectively. The problem 
defined in 
Eq.~\eqref{eq-prob-def-c} has some properties that 
will be useful to investigate. The function which we maximize 
is convex with respect to the initial state $\rho$ as well as it is convex 
with respect to each operation $a, b, c$ applied. Hence, we should choose 
$\rho$ as a deterministic bit, w.l.o.g. we take $\rho = 0$. We should choose 
stochastic operations $a, b, c$ as a deterministic as well. For each 
composition order $\pi$ the output $\pi(a,b,c)(\rho)$ is $0$ or $1$. 
As the outputs are deterministic, then the 
optimal measurement $\{Q_\pi\}_\pi$ counts the number of different 
outputs. From this we see that the optimal probability will not exceed 
$1/3$. We can easily achieve this value by considering $a(x) = 1, b(x) 
= 1 - 
x$ and $c(x) = x$, for $x \in \{0, 1\}$. It implies that $cba(0) = 0$ and 
$acb(0) 
= 1$. In Fig. \ref{fig:cb} we present a sketch of the optimal strategy with 
two 
exemplary orders. 
We conclude the optimal probability of the compositions' order  
discrimination is given by 
\begin{equation}
p_{\mathrm{c}}  = \frac{1}{3}.
\end{equation}

\subsection{\bf Classical strategy based on local operations and shared 
randomness:}\label{losr}

\begin{figure}[htp!]
	\centering
		\includegraphics[width=0.55\textwidth]{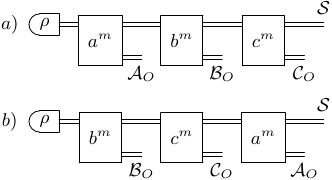}
	\caption{A graphical representation of possible orders of the 
	bi-partite mappings $a^m, b^m, c^m$ with the initial 
	state $\rho$. Double lines represent a classical information. Time flows 
	from left to right. The  order of operations  a) $a^m \to b^m \to c^m$; b) 
	$b^m 
	\to  c^m \to 
	a^m$.}
	\label{fig:random}
\end{figure}

In this section, we extend the classical memoryless approach to a strategy 
based on local operations and shared randomness, where parties are allowed to 
save some information on their local auxiliary systems. In this scenario, the 
actions of Alice, Bob, and Charlie are given by classical, bi-partite 
stochastic maps $a^m 
\in 
\mathrm{L}\left(\A_I \otimes \mathcal{S}, \A_O \otimes \mathcal{S}\right)$, 
$b^m \in 
\mathrm{L}\left(\B_I \otimes \mathcal{S}, \B_O \otimes \mathcal{S}\right)$ 
and $c^m \in 
\mathrm{L}\left(\CC_I \otimes \mathcal{S}, \CC_O \otimes 
\mathcal{S}\right)$, 
respectively. The dimension of the shared system $\SS$ remains $\dim S = 
2$, but $\A_I, \A_O, \B_I, \B_O, \CC_I, \CC_O$ are arbitrary, 
finite-dimensional real vector spaces. Similarly, as for classical memoryless 
strategies, we may exploit the convexity of the problem 
defined in Eq.~ \eqref{eq-prob-def-c} to find the form 
of optimal initial state $\rho \in \Omega\left(\SS 
\otimes \A_I \otimes \B_I \otimes \CC_I\right)$, linear mappings $a^m, b^m, 
c^m$ and the final 
measurement $\{Q_\pi\}_\pi$. The initial state $\rho$ should be  again 
deterministic. That means $\rho$ is a product state, which implies there is 
no need to share randomness among all parties. Hence, we may fix 
$\rho$ to be defined on the space $\SS$, w.l.o.g. as $\rho = 0$. Meanwhile, 
the classical 
operations are deterministic as well and now, they are of the form: $a^m \in 
\mathrm{L}\left(\mathcal{S}, \A_O \otimes \mathcal{S}\right)$, 
$b^m \in 
\mathrm{L}\left(\mathcal{S}, \B_O \otimes \mathcal{S}\right)$ 
and $c^m \in 
\mathrm{L}\left(\mathcal{S}, \CC_O \otimes 
\mathcal{S}\right)$. Next, taking as an example $a^m$, for any $x \in \{0, 
1\}$ we get $a^m(x) = \left(a^m(x)_1, a^m(x)_2\right)$, where $a^m(x)_1$ is 
defined on $\SS$ and $a^m(x)_2$ is defined on $\A_O$. It is 
enough to consider $a^m(x)_2 = x$ and carry the function $x \mapsto 
a^m(x)_2$ to the final measurement. Hence, we restrict our attention to the 
operation of the form $a^m(x) = \left(a^m(x)_1, x\right) \in \SS \otimes 
\A_O$, where $\dim \A_O = 2$. We apply the same for $b^m$ and $c^m$.
A schematic representation of this strategy is presented in 
Fig.\ref{fig:random}.

Let us now identify $a^m$ by the tuple $(a^m(0)_1, a^m(1)_1) \in \{0, 1\}^{2}$ 
and 
similarly for $b^m, c^m$. We clearly see that each of the parties has $4$ 
possible operations to consider. Hence, we have in total $64$ strategies in 
total to investigate to find the optimal one. For any $\pi \in S_3$, the 
output $\pi(a^m, b^m, c^m)(0)$ is a deterministic 
tuple of the form $(x, x_a, x_b, x_c)$, where $x$ is the output on $\SS$ and 
$x_a, x_b, x_c$ are the values of the inputs that Alice, Bob and Charlie 
received. The optimal measurement  $\{Q_\pi\}_\pi$ counts the number of 
different output tuples that are achievable, and therefore the probability of 
success will be equal to $\frac{1}{6} \left| \{\pi(a^m, b^m, c^m)(0) : \pi 
\in S_3 \} \right|$. The code performing these calculations is available in 
repository~\cite{code22}. For $\rho = 0$ the optimal strategy 
is of the form $a^m = (0, 0)$, $b^m = c^m = (1, 1)$. It returns 
$5$ unique tuples: $cba(0) = (1, 0, 0, 1), bca(0) = (1, 0, 1, 0), cab(0) = 
bac(0) = (1, 1, 0, 0), acb(0) = (0, 1, 0, 1),$ and $abc(0) = (0, 1, 1, 0)$. 
Therefore, in that case,  we get
\begin{equation}
p_c^m = \frac{5}{6}.
\end{equation}

\subsection{\bf Classical non-signaling strategy:}\label{nonsign}
\begin{figure}
\includegraphics[width=0.4\textwidth]{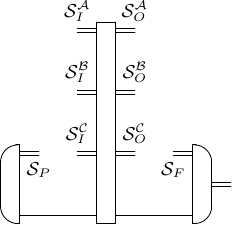}
\caption{A schematic representation of a general discrimination strategy involving classical 
non-signaling 
operations. Double lines represent a classical information. Time flows 
	from left to right.}
\label{fig:nons}
\end{figure}

The last classical strategies we investigate are based on non-signaling 
operations. They represent the upper limit for capabilities of all classical 
strategies. The notation is based on the formalism of process 
matrices~\cite{oreshkov2012quantum}, 
quantum networks~\cite{chiribella2009theoretical} and non-signaling quantum 
channels~\cite{piani2006properties} - 
typically used for the analysis of quantum systems. The classical solutions 
are submerged within that notation. Generally, the order of the 
operations' composition can be formally 
represented by using process matrices~\cite{oreshkov2012quantum, 
araujo2015witnessing} in the following way. Let us 
introduce the notation of qubit spaces $\C^2$: $\SS_P, \SS_F, \SS_I^{\A}, 
\SS_O^{\A}, 
\SS_I^{\B}, \SS_O^{\B},\SS_I^{\CC}, \SS_O^{\CC}$, where $\SS_P$ is the space 
of the input quantum state, $\SS_F$ is the space of the quantum measurement 
input, $\SS_I^{\A}$ 
is the space of Alice's system input and so on (see Fig.~\ref{fig:nons}). 
For a given permutation $\pi \in S_3$, we 
introduce a permutation process matrix $
W_\pi = \projV{\Id}_{\SS_P, \SS_I^{\pi(1)}} \otimes 
\projV{\Id}_{\SS_O^{\pi(1)}, \SS_I^{\pi(2)}} \otimes 
\projV{\Id}_{\SS_O^{\pi(2)}, \SS_I^{\pi(3)}} \otimes 
\projV{\Id}_{\SS_O^{\pi(3)}, \SS_F}$, where $\pi(i) \in \{\A, \B, \CC\}$ for 
$i=1,2,3$. The introduced process matrices $W_\pi$ represent the unknown order 
parametrized by the permutation $\pi$. From the duality of process matrices 
\cite{milz2021resource}, the most 
general scheme that can be 
used to discriminate causal order is based on non-signaling operations given 
in the form of quantum network (see 
Fig.~\ref{fig:nons}). Let
$\OO = \C^{6}$ represents the measurement output space and let 
 $D = \sum_{\pi \in S_3} \proj{\pi}_\OO \otimes 
D_\pi $ stands for the Choi-Jamio{\l}kowski 
representation~\cite{choi1975completely, 
jamiolkowski1972linear} of a 
discrimination strategy defined on the space $\SS_P \otimes 
\SS_I^\A \otimes \SS_O^\A \otimes \SS_I^\B \otimes \SS_O^\B \otimes 
\SS_I^\CC \otimes \SS_O^\CC \otimes \SS_F \otimes 
\OO$. The operator $D$ satisfies couple of conditions. First, it is positive 
semidefinite. Second, it 
represents a network with an input state, non-signaling 
channel~\cite{beckman2001causal, 
piani2006properties} in the middle and the measurement at the end, which is 
mathematically defined as: $\tr_{\OO}(D) = \Id_{\SS_F}/2 \otimes \tr_{\OO, 
\SS_F}(D)$, $\tr_{\OO, \SS_O^\A}(D) = \Id_{\SS_I^\A}/2 \otimes \tr_{\OO, 
\SS_O^\A, \SS_I^\A}(D)$, $\tr_{\OO, \SS_O^\B}(D) = \Id_{\SS_I^\B}/2 \otimes 
\tr_{\OO, 
\SS_O^\B, \SS_I^\B}(D)$, $\tr_{\OO, \SS_O^\CC}(D) = \Id_{\SS_I^\CC}/2 
\otimes 
\tr_{\OO, 
\SS_O^\CC, \SS_I^\CC}(D)$ and $\tr(D) = 16$, where $\tr_{\XX}$ stands for the 
partial 
trace over the subsystem $\XX$~\cite{watrous2018theory}. 

As the considered 
strategy is 
classical, our network has to take into account that condition. All inputs and 
outputs has to accept only classical states, however, inside the network it is 
allowed to create some non-trivial correlations by quantum effects. Hence, we 
have to add the final condition $\Delta(D) = D$, where $\Delta$ is the 
completely dephasing channel, which maps
any operator into its diagonal form. The concatenation of the strategy $D$ 
with 
the order permutation process 
matrix $W_\pi$ provides the distribution of the outputs $\pi$ defined on 
$\OO$. Formally this concatenation is given by a link 
product~\cite{chiribella2009theoretical}, 
$\tr_{\not\OO}\left(D 
(\Id_\OO 
\otimes W_\pi^\top) \right)$, where $\tr_{\not\OO}$ indicates the trace of 
over all subsystems except $\OO$. Given $\pi$ is the actual order, the 
probability we guessed it correctly equals $\bra{\pi}\tr_{\not\OO}\left(D 
(\Id_\OO 
\otimes W_\pi^\top) \right) \ket{\pi} = \tr(D_\pi W_\pi)$. The total 
probability of success reads then $\frac{1}{6} \sum_{\pi \in S_3} \tr 
(W_{\pi}D_\pi)$. We can write the optimization problem 
to calculate the optimal probability of success. As the objective function 
is linear in $D$ and presented constraints are linear and conic, we will end 
up 
with SDP optimization presented in Program~\ref{sdp-non}. 
\begin{table}[!htp]
	\begin{center}
		\centering\underline{Primal problem}
		\begin{equation*}
		\begin{split}
		\text{maximize:}\quad &
		\frac{1}{6} \sum_{\pi \in S_3} \tr (W_{\pi}D_\pi)
		\\[2mm]
		\text{subject to:}\quad & 
		\sum_{\pi \in S_3} D_\pi =  \frac{\Id_{\SS_F}}{2} \otimes \sum_{\pi \in S_3} 
		\tr_{\SS_F}(D_\pi),  \\ 
& 	\sum_{\pi \in S_3} \tr_{\SS_F, \SS_O^\A}(D_\pi)  = 
\frac{\Id_{\SS_I^\A}}{2} 
	\otimes 
	\sum_{\pi \in S_3} \tr_{\SS_F, \SS_I^\A, \SS_O^\A}(D_\pi),\\
& 	\sum_{\pi \in S_3} \tr_{\SS_F, \SS_O^\B}(D_\pi)  = 
\frac{\Id_{\SS_{I}^\B}}{2}
	\otimes 
	\sum_{\pi \in S_3} \tr_{\SS_F, \SS_I^\B, \SS_O^\B}(D_\pi),\\
& 	\sum_{\pi \in S_3} \tr_{\SS_F, \SS_O^\CC}(D_\pi)  = 
\frac{\Id_{\SS_{I}^\CC}}{2} 
	\otimes 
	\sum_{\pi \in S_3} \tr_{\SS_F, \SS_I^\CC, \SS_O^\CC}(D_\pi),\\
 & \sum_{\pi \in S_3}\tr\left( 
		D_\pi\right) = 16,\\& D_\pi = \Delta(D_\pi),\\
		&  
		D_\pi \in \mathrm{Pos}\left(\SS_P \otimes 
\SS_I^\A \otimes \SS_O^\A \otimes \SS_I^\B \otimes \SS_O^\B \otimes 
\SS_I^\CC \otimes \SS_O^\CC \otimes \SS_F\right),\\&\pi \in S_3.
		\end{split}
		\end{equation*}
	\end{center}
\caption{SDP program for computing the maximum value of the probability of successful 
	order discrimination using classical non-signaling strategies. 
	\label{sdp-non}}
\end{table}  
To optimize this problem, we use \texttt{Julia}
programming language along with quantum package 
\texttt{QuantumInformation.jl} \cite{Gawron2018} and
SDP optimization via SCS solver \cite{ocpb:16, scs} with absolute 
convergence tolerance $10^{-8}$. The
code is available on GitHub \cite{code22}. Finally, the optimal probability 
which can 
be achieved in classical non-signaling scenario is
\begin{equation}
p_c^{ns} = \frac{5}{6}.
\end{equation}

The remainder of this section is devoted to analyzing the quantum case.

\subsection{\bf Quantum memoryless strategy:}\label{quantum-basic}
\begin{figure}[htp!]
	\centering
	\includegraphics[scale=0.85]{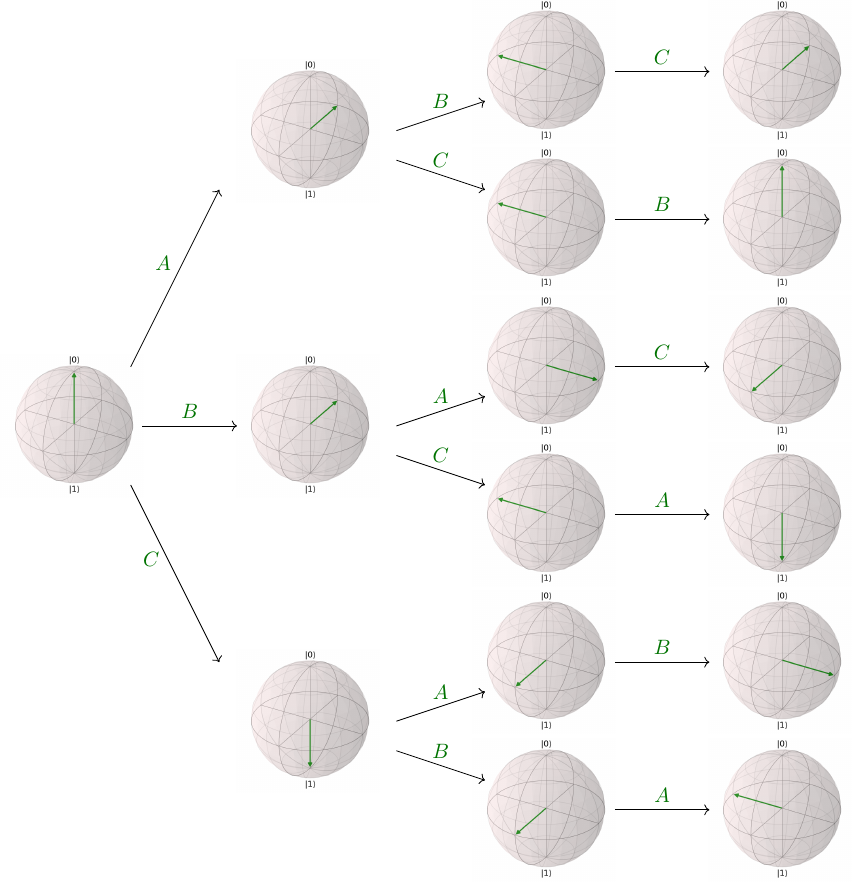}
	\caption{A schematic representation of the 
	construction of three mutually unbiased bases $B_{\Id}, B_{H}$ and 
	$B_{\widetilde{H}}$ on the Bloch sphere. The orders from the top to 
	botton are: $CBA\ket{0} \simeq \ket{-}, BCA\ket{0} \simeq \ket{0}, 
	CAB\ket{0} \simeq \ket{+}, ACB\ket{0} \simeq \ket{1}, BAC\ket{0} \simeq 
	\ket{i}, ABC\ket{0} \simeq \ket{-i}$. }
	\label{bloch2}
\end{figure}

In this scenario Alice, Bob and Charlie are using quantum, mono-partite 
channels: $A, B, C \in \mathrm{L}(\mathcal{S})$. For any quantum state 
$\rho \in 
\Omega(\SS)$ and permutation $\pi \in S_3$ it holds $\pi(A,B,C)(\rho) \le 
\Id_\SS$. We obtain the 
	following upper bound for the Eq.~\eqref{eq-prob-def}: $
	\frac{1}{6} \sum_{\pi} \tr \left( Q_\pi 
	\pi(A,B,C)(\rho) \right) \le \frac{1}{6} \sum_{\pi} \tr 
	\left( Q_\pi \right) = \frac{1}{3}$. That implies quantum memoryless 
	strategies do not improve the probability of composition 
order discrimination over classical ones, which means
	\begin{equation}
 	p_{\mathrm{q}}  = \frac{1}{3}.
 	\end{equation}
 However, there exists a particular choice of quantum channels $A, B, C$ that 
 returns a set of the most separated quantum states, creating three mutually 
 unbiased qubit bases~\cite{durt2010mutually}. In particular we define three 
 unitary operations given by their Kraus operator: $A = \frac{1}{\sqrt{2}}
 \left(\begin{array}{cc} 1&-i
\\-1&-i\end{array}\right),
B =\frac{1}{\sqrt{2}} \left(\begin{array}{cc}1&i 
\\-1&i\end{array}\right), C =  \frac{1}{\sqrt{2}}
\left(\begin{array}{cc}0 & 1-i 
\\ 1+i &0 \end{array}\right)$.
Then,  all compositions of the given operators starting from $\rho = \proj{0}$ 
create three 
mutually unbiased bases $B_\Id, B_H$ and $B_{\widetilde{H}}$ in 
$\C^2$  given by $
B_\Id = \left\{ \ket{0}, \ket{1} \right\}$,\\
$B_H = \left\{ \ket{+}\coloneqq\frac{1}{\sqrt{2}}(\ket{0}+\ket{1}),
\ket{-}\coloneqq\frac{1}{\sqrt{2}}(\ket{0}-\ket{1})\right\}$ and \\ 
$B_{\widetilde{H}} = 
\left\{ \ket{i} \coloneqq \frac{1}{\sqrt{2}}(\ket{0}+i\ket{1}), 
\ket{-i} \coloneqq \frac{1}{\sqrt{2}}(\ket{0}-i\ket{1}) \right\}.$
A schematic representation of this construction and output states analysis 
is presented on Bloch 
sphere in Fig.\ref{bloch2}. 

\subsection{\bf Quantum strategy based on local operations and shared 
entanglement:}\label{lose}
\begin{figure}[htp!]
	\centering
		\includegraphics[width=0.75\textwidth]{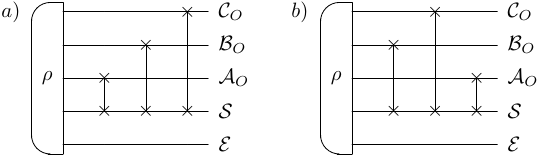}
	\caption{A graphical representation of a discrimination strategy 
	involving quantum shared entanglement. All parties use bi-partite swap 
	$U$ acting on $\SS$ and an appropriate input space. Time flows 
		from left to right. Observe different order of swap gates presented in 
		circuits a) $A^m \to B^m \to C^m$; b) 
	$B^m 
	\to  C^m \to 
	A^m$. }
	\label{fig:lose}
\end{figure}

In this section we will present an advantage of using quantum strategies 
utilizing shared entanglement. Below we present the optimal strategy for 
which we achieve perfect discrimination of causal orders. In this scenario 
Alice, Bob and Charlie share a quantum state 
$\rho \in \Omega(\A_I \otimes \B_I \otimes \CC_I \otimes \SS \otimes 
\mathcal{E}) $, 
where $\mathcal{E}$ is an auxiliary system that parties will have access at 
the measurement stage. Here, dimensions of $\A_I, \B_I, \CC_I$ and 
$\mathcal{E}$ are arbitrary. The parties are 
using quantum, bi-partite channels $A^m \in 
\mathrm{L}\left(\A_I \otimes \mathcal{S}, \A_O \otimes \mathcal{S}\right)$, 
$B^m \in 
\mathrm{L}\left(\B_I \otimes \mathcal{S}, \B_O \otimes \mathcal{S}\right)$ 
and $C^m \in 
\mathrm{L}\left(\CC_I \otimes \mathcal{S}, \CC_O \otimes 
\mathcal{S}\right)$. From the convexity of the problem 
Eq.~\eqref{eq-prob-def} the initial state should be a pure state and by 
the Schmidt decomposition~\cite{watrous2018theory} it is 
enough to consider $\EE \simeq \A_I \otimes \B_I \otimes \CC_I \otimes \SS$. 
Also, as $\A_I, \B_I, \CC_I$ are arbitrary by a Stinespring representation of 
quantum channels~\cite{watrous2018theory} it is enough to consider unitary 
quantum channels and hence, assume that $\A_I \simeq \A_O, \B_I \simeq \B_O$ 
and 
$\CC_I 
\simeq \CC_O$. In our construction we restrict dimensions of the subsystems 
to have values $\dim \A_I = \dim \A_O = \dim \B_I = \dim \B_O = \dim \CC_I = 
\dim 
\CC_O = 2$. It is assumed that each party uses the same operation - a swap 
between system $\SS$ and an appropriate 
input space given by the unitary matrix $U = \sum_{i, j} \ketbra{i,j}{j,i}$ 
. The input state $\rho = \projV{\sqrt{\sigma}}$ is taken as a 
purification of the state $\sigma \in \Omega(\A_I \otimes \B_I \otimes \CC_I 
\otimes \SS)$. For each $\pi \in S_3$ the composition $\pi(A^m, B^m, C^m)$ is 
represented by unitary matrix $\pi(U, U, U) = \left(U_{\SS, \pi(3)_I}U_{\SS, 
\pi(2)_I}U_{\SS, 
\pi(1)_I}\right) \otimes \Id_\EE$, where $\pi(i) \in \{\A, \B, \CC\}$ for 
$i=1,2,3$. Let us denote by $M_\pi = U_{\SS, \pi(3)_I}U_{\SS, \pi(2)_I}U_{\SS, 
\pi(1)_I}$ a system permutation matrix for $\pi$, e.g. if Alice is before Bob 
and Bob is 
before Charlie, 
then $M_\pi = U_{\SS, \CC_I} U_{\SS, \B_I} U_{\SS, \A_I}$ (see 
Fig.~\ref{fig:lose}).
To achieve perfect 
discrimination in this scenario, we need to check if there exists a state 
$\sigma$ for which the output pure states (vectors) $\pi(U, U, 
U)\ketV{\sqrt{\sigma}} = (M_\pi \otimes \Id_\EE) 
\ketV{\sqrt{\sigma}} = \ketV{M_\pi \sqrt{\sigma}}$ 
are orthogonal or simplifying, for which it holds 
$\tr\left(M^\dagger_{\pi'}M_\pi \sigma \right) = 0$ for $\pi \neq 
\pi'$. We find such $\sigma$ by using Semidefinite programming (SDP) 
optimization 
procedure~\cite{code22} presented in Program~\ref{sdp-rho}. 
\begin{table}[!h]
	\begin{center}
		\centering\underline{Primal problem}
		\begin{equation*}
		\begin{split}
		\text{maximize:}\quad &
		\tr (\sigma)
		\\[2mm]
		\text{subject to:}\quad & 
		\tr \left( M^\dagger_{\pi'}M_\pi \sigma \right) = 0,\\& 
		\pi' \neq \pi, \\ & 
		\pi, \pi' \in S_3,\\& 
		\sigma \in \mathrm{Pos}(\A_I \otimes \B_I \otimes \CC_I 
\otimes \SS), \\ & 
		\tr(\sigma) \le 1. 
		\end{split}
		\end{equation*}
		\caption{Convex program that searches if there is an initial state 
		$\rho = \projV{\sqrt{\sigma}}$ that guarantees perfect 
		discrimination of composition order for quantum strategies based on 
		local operations and shared entanglement.}
		\label{sdp-rho}
	\end{center}
\end{table}
The program returns a numerical approximation (with tolerance $10^{-8}$) of 
the state
\begin{equation}
\sigma = \frac{1}{12}\Id - \frac{1}{15}\left(\tau_0 +\tau_1 + 
\tau_2 + \tau_3 + \tau_4\right),
\end{equation} 
where $\tau_i = \proj{x_i}$ and $\ket{x_i} \coloneqq \ket{D_4^i}$ is an $i$-th 
Dicke quantum state defined on $4$ qubits~\cite{dicke1954coherence, 
burchardt2021entanglement}, 
for example: $\ket{x_1} = \frac{1}{2}\left(\ket{0,0,0,1} + 
\ket{0,0,1,0} + 
\ket{0,1,0,0} + \ket{1,0,0,0}\right)$ also called $\ket{W_4}$ state, while 
$\ket{x_2} = \frac{1}{\sqrt{6}}(\ket{0, 0, 1, 1} + \ket{0, 1, 0, 1} + 
\ket{1, 0, 0, 1} + \ket{0, 1, 1, 0} + \ket{1, 0, 1, 0} + \ket{1, 1, 0, 0})$. 
We can check 
directly that $\sigma$ is well defined quantum state, that is $\sigma \ge 0$ 
and $\tr(\sigma) = 1$. Moreover, for any system permutation matrix $M$ we 
have  
$\tr\left(M \sigma \right) = \frac{1}{12}\tr(M) - \frac{1}{3}$. In 
particular, if  $M = M^\dagger_{\pi'}M_\pi$ for $\pi\neq \pi'$ and $\pi, 
\pi' \in S_3$, we have $\tr\left( M^\dagger_{\pi'}M_\pi \sigma \right) = 0$ 
(we provide additionally a tool that 
checks $\tr\left( M^\dagger_{\pi'}M_\pi \right) = 4$ with no error 
tolerance~\cite{code22}). To summarize, we confirmed 
that using quantum shared 
entanglement gives the probability of 
successful discrimination equal to
\begin{equation}
p_q^{m} = 1.
\end{equation}

 \section{Conclusion and discussion} \label{sec:conclusion}

In this work, we studied an explicit three-party
communication task for which players Alice, Bob, and Charlie meet to discuss 
a strategy such that after separation, they would discover the 
hidden order in which their moves are made. 
We have considered both classical and quantum scenarios. We proved that for 
both classical and quantum memoryless strategies the optimal probability of  
order discrimination equals $p_c = p_q = 1/3$. It is worth mentioning, we 
showed that it is possible to find three single-qubit unitary operations, such 
that acting with them on a selected state in different order one obtains three 
pairs of orthogonal states, which jointly form the set of three mutually 
unbiased qubit bases. Next, we have extended 
the classical memoryless scenario to the strategy based on local operations 
and shared randomness and the non-signaling strategy. In both scenarios, we 
improved the success probability to $p_c^m=p_c^{ns} = 5/6$. In 
particular, we showed with the assistance of numerical optimization that 
classical strategy based on local operations with additional bit memory for 
each party is optimal among all classical strategies. 
Next, we considered the quantum LOSE approach for the order discrimination. We 
have shown an advantage of this scenario, which gave us perfect 
discrimination, $p_q^m = 1$.

This work paves the way toward a complete description of the capabilities of 
quantum advantage in the communication task of causal order discrimination 
between parties. The next step of the research could be focused on $N$-party 
scenario, where the dimension of shared system $\SS$ is arbitrary, 
$\dim \SS = d$. We would like to calculate exact value of the function $(d, 
N) \mapsto (p_c, p_q, p_c^m, p_q^m, p_c^{ns}, p_q^{ns})$, where $p_q^{ns}$ 
represents the probability for quantum scenario with non-signaling operations.
 Alternatively, the analysis of the asymptotic behavior of the probability 
 function with fixed $d$ and  $N \rightarrow \infty$ could be interesting, yet 
 easier to handle. Additionally, one could use multiple rounds of the same 
 order to increase the probability of success and to investigate the optimal 
 strategies as well as the probability of success in that case.  
 
\section*{Acknowledgements}

RK and KŻ acknowledge financial support by the National Science Centre, Poland, 
under the contract 
number 2021/03/Y/ST2/00193 within the QuantERA II Programme that has 
received funding from the European Union’s Horizon 2020 research and 
innovation programme under Grant Agreement No 101017733.

PL acknowledges financial support by the Ministry of Education, Youth and Sports of the Czech Republic through the e-INFRA CZ (ID:90254),
with the financial support of the European Union under the REFRESH – Research Excellence For REgionSustainability and High-tech Industries project number CZ.10.03.01/00/22\_003/0000048 via the Operational Programme Just Transition.

\bibliographystyle{unsrt}
\bibliography{switch.bib}

\end{document}